# SPEECH EMOTION RECOGNITION USING SELF-SUPERVISED FEATURES

*Edmilson Morais, Ron Hoory, Weizhong Zhu, Itai Gat, Matheus Damasceno and Hagai Aronowitz*

IBM Research AI

## ABSTRACT

Self-supervised pre-trained features have consistently delivered state-of-art results in the field of natural language processing (NLP); however, their merits in the field of speech emotion recognition (SER) still need further investigation. In this paper we introduce a modular End-to-End (E2E) SER system based on an Upstream + Downstream architecture paradigm, which allows easy use/integration of a large variety of self-supervised features. Several SER experiments for predicting categorical emotion classes from the IEMOCAP dataset are performed. These experiments investigate interactions among fine-tuning of self-supervised feature models, aggregation of frame-level features into utterance-level features and back-end classification networks. The proposed monomodal speech-only based system not only achieves SOTA results, but also brings light to the possibility of powerful and well fine-tuned self-supervised acoustic features that reach results similar to the results achieved by SOTA multimodal systems using both Speech and Text modalities.

*Index Terms*— Speech emotion recognition, self-supervised features, end-to-end systems.

## 1. INTRODUCTION

Speech emotion recognition (SER) is an essential capability for Human Computer Interaction (HCI) and has attracted much attention due to its wide range of potential application in e.g. call center conversation analysis, mental health and spoken dialogue systems. Although significant progress has been made [1], SER is still a challenging research problem since human emotions are inherently complex, ambiguous, and highly personal. Humans often express their emotions using multiple simultaneous cues, such as voice characteristics, linguistic content, facial expression, and body actions, which makes SER by nature a complex multimodal task [2]. In addition to that, due to the difficulties in data collection, publicly available datasets often do not have enough speakers to properly cover personal variations in emotion expression. As a consequence of that, some of the most common Deep Learning techniques that have been incorporated to SER are related to the fields of: transfer learning [3, 4]; multitask learning [5];

multimodal systems [6, 7], and more powerful model architecture [8, 9]. Despite the success of self-supervised features in the field of natural language processing (NLP) in the last year and, more recently, in speech recognition and speaker identification tasks, only few works have investigated this kind of features in the context of SER [5, 10, 11]. Therefore, we argue that the merits of self-supervised features in the field of SER still need further investigation. Moreover, in applications where speech is the only modality available, such as in call center conversation analysis, the other modalities such as facial expression and body actions are unusable and, therefore, the SER system must rely only on the information available on the speech signal.

Given this scenario, the main goals of this paper are: (1) to introduce a modular End-to-End (E2E) SER system based on an Upstream + Downstream architecture model paradigm which allows easy use/integration of a large variety of self-supervised features, (2) to present a sequence of experiments comparing/analyzing the performance of the proposed E2E system under different configurations and (3) to show that, despite using only the speech modality, the proposed E2E system can reach SOTA results compared to the SOTA results achieved by multimodal systems which use both Speech and Text modalities.

The rest of this paper is organized as follow: Session 2 presents the proposed End-to-End SER model; The experimental setup is presented in Section 3. Section 4 presents the results. Finally, conclusions and future works are presented in Section 5

## 2. PROPOSED MODEL

In this paper we formulate the SER problem as a mapping from the continuous speech domain into the discrete domain of categorical labels of emotion. The paradigm used here is the Upstream + Downstream Model which is illustrated in Figure 1.

This Upstream + Downstream model is very similar to what has largely been used by the NLP community since the introduction of BERT [12] and it has now become more popular for the speech community with the advent of works like Wav2vec 2.0 [13, 14], TERA [15], Mockingjay [16], huBERT [17] and projects like SUPERB [18] and SpeechBrain [19].

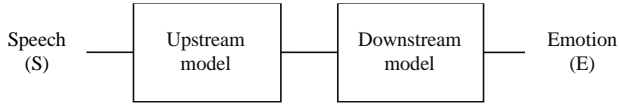

Figure 1. Proposed SER model architecture

In our SER model, the Upstream is a task-independent model, it is pre-trained in self-supervised fashion and it works as an Encoder or Front-End model responsible for feature extraction; on the other hand, the Downstream is a task-dependent model, it works as a Decoder or Back-End model and it is responsible for the final task of classifying the features generated by the Upstream model into categorical labels of emotion.

The Self-Supervised Upstream Models used in this work are the latest release of Wav2Vec 2.0 [14] and the Hidden Unit BERT (huBERT) [17]. The Downstream models used in this work are: (1) A simple Mean Average Pooling aggregator followed by a linear classifier and (2) the ECAPA-TDNN aggregator [20] followed by a linear classifier.

### 2.1. Dataset

The IEMOCAP [21] corpus used in this paper is a multimodal dyadic conversational dataset. It consists of approximately 12 hours of multimodal data, including speech, text transcriptions and facial recordings. IEMOCAP contains a total of 5 sessions and 10 different speakers, with a session being a conversation of two exclusive speakers. To be consistent and to be able to compare with previous studies, only utterances with ground truth labels belonging to "angry", "happy", "excited, "sad", and "neutral" were used. The "excited" class was merged with "happy" to better balance the size of each emotion class, which results in a total of 5,531 utterances (happy 1636, angry 1,103, sad 1,084, neutral 1,708). Unless otherwise stated, leave-one-session-out 5-fold cross validation (CV) is used and the average result reported. At each fold of the 5-fold CV set-up, 2 speakers are used for testing and the samples from the 8 speakers remaining are randomly split into 80% for training and 20% for validation. The splitting done here is exactly the same as the one done by SUPERB [18], which splits each of the 5 IEMOCAP folds into three subsets: a training-set, a validation-set and a test-set.

### 2.2. Fine-tuning of the upstream models

In order to boost the performance of our SER system we fine-tuned our Upstream models using the categorical labels of emotion from the IEMOCAP dataset described in Section 2.1. The fine-tuning of our Upstream model is performed by training it jointly with a simple Mean-Average Pooling Network followed by a Linear Classifier, as described in Figure 2.

This fine-tuning is performed for each of the 5-folds of IEMOCAP dataset. Therefore, at the end of the fine-tuning process we come up with 5 different fine-tuned Upstream models (one for each of the 5-folds).

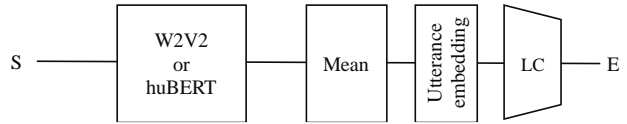

Figure 2. Upstream model fine-tuning process

### 2.3. Average of checkpoints

In addition to the process of fine-tuning, in order to minimize the output variance of the Upstream models, the 5 best fine-tuned Upstream Model checkpoints are averaged [22]. This processing is done separately for each of the 5-folds from the IEMOCAP dataset. The selection of the 5 best checkpoints to be averaged (for each of the 5-folds) is done solely based on the accuracy of the fine-tuned Upstream Model on the validation-set. The test-set is not observed by this process at any moment.

This procedure of fine-tuning (FT) followed by Checkpoint Averaging (AVG) has been performed for both Wav2Vec 2.0 and huBERT, generating a total of 10 fine-tuned and averaged Upstream Models: 5 FT-AVG Wav2Vec 2.0 and 5 FT-AVG huBERT models.

Given these fine-tuned and averaged Upstream Models for Wav2Vec 2.0 and huBERT and the two possible variations of our Downstream model we have conducted several experiments which are described in Figure 3, Figure 4 and in Table 1.

## 3. EXPERIMENTAL SETUP

### 3.1. Experiments for evaluation

Among the several experiments that can be performed using our SER model, we have selected the experiments illustrated in Figure 3 and Table 1 in the Sets (1.A) and (1.B). The goal of these experiments is to understand the following: (1) the importance of fine-tuning the Upstream model; (2) the importance of averaging the Upstream and Downstream Model Checkpoints; (3) how Wav2vec 2.0 and huBERT can be combined to boost SER performance and (4) the performance of the two aggregators used: Mean Pooling and ECAPA-TDNN. Unless otherwise stated, in all following experiments both Upstream and Downstream model checkpoints are averaged following the same procedure described in Section 2.3 [22].

Referring to Figure 3, in experiments (1-4) the Upstream model used is either Wav2vec 2.0 or huBERT and the Downstream model is composed by a Mean Average Aggregator followed by a linear classifier. In experiments in (1-2) neither Upstream nor Downstream model are averaged at all. In experiments in (3-4) both Upstream and Downstream models are averaged.

Experiments (5-6) are similar to experiments (3-4), the only difference is the Aggregator used, which is now the ECAPA-TDNN.

In the experiment (7) we have an early fusion between the Wav2vec 2.0 and the huBERT features, which is done just before going through the ECAPA-TDNN aggregator.

In the experiment (8) we have a later fusion between the utterance embeddings generated by two ECAPA-TDNNs, the first one operating on the Wav2vec 2.0 features and the second one operating on the huBERT features.

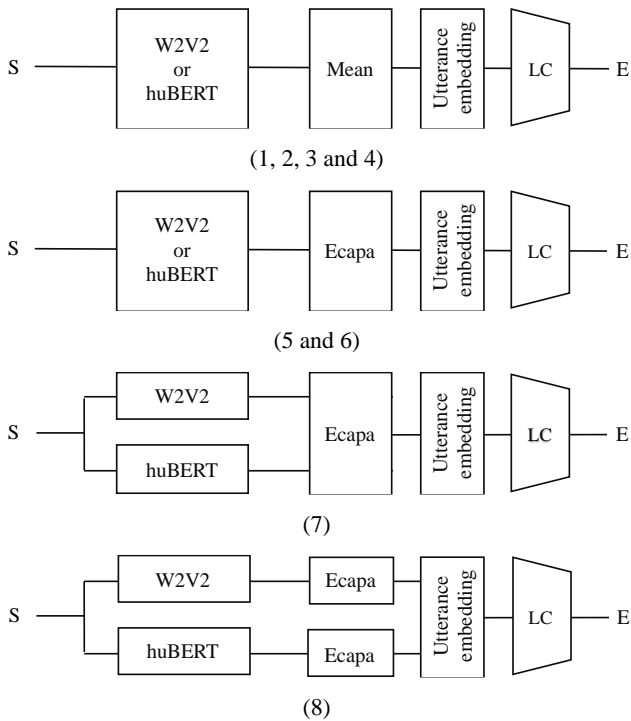

Figure 3. Experiments selected to evaluate our SER model.

### 3.2. Experiments to be used as baselines

In addition to the experiments from Figure 3 we have also prepared the baseline/reference experiments presented in Figure 4. These baseline models use standard hand-crafted Filter-Bank features for the acoustic modality and a BERT model to extract feature embeddings for the text modality. This BERT model has also been fine-tuned and averaged using the same processes describe in Sections 2.2 and 2.3. It is important to emphasize that the Fbank used here does not have explicit pitch information attached to it and that the fine-tuning optimization process of the BERT model may not follow the most advanced SOTA techniques available nowadays. However, despite not being as carefully prepared as it could be, these baseline models can help us to obtain insight on how powerful these fine-tuned and averaged Wav2vec 2.0 and huBERT features are.

Referring to Figure 4, in experiment (9) a standard Filter-bank is used as the Upstream Model. In experiment (10) BERT model is used as the Upstream model. Finally, in experiment (11) a standard Fbank for the acoustic modality and BERT features for the text modality are used jointly in a later fusion fashion.

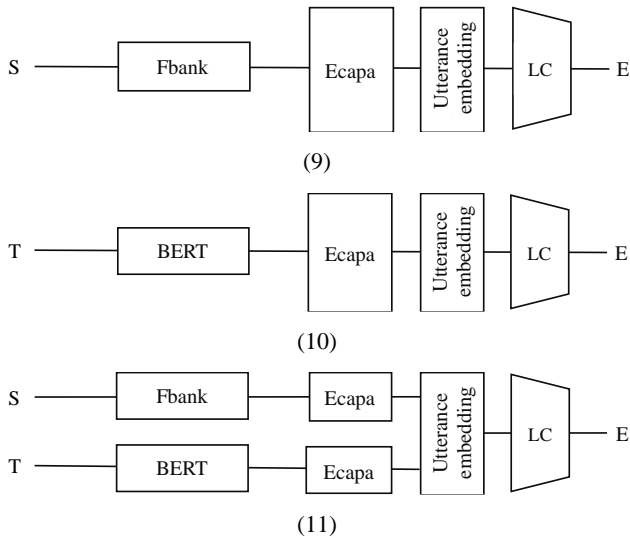

Figure 4. Experiments used as baseline/reference models to be compared with the experiments in Figure 3.

In addition to the baselines in Figure 4 a carefully comparison between the results achieved by our SER model and the SOTA results available on the literature will be done in Section 4.1.

## 4. RESULTS

Results of the experiments presented in Figure 3 and 4 are summarized in Table 1 and described in detail bellow:

In column 2 of Table 1, under the term (#), we indicate the number of the 11 experiments evaluated. In column 3 we indicate the input modality used in each experiment.

In column 4 under the term **Upstream model** we can find the indication of the **Input feature**; if the Upstream model has been fine-tuned (**FT**); and if the Upstream model has been Averaged (**AVG**). The symbol "+" in experiment 7 (huBERT + W2V2) indicates early fusion of the features and the symbol "&" in the experiment 8 and 11 indicates later fusion of the features.

In column 5 under the term **Downstream model** we can find the indication of the Aggregator Model used (**AGG**); the Classification Model (**Classifier**) used; and if the full Downstream Model has been averaged (**AVG**).

Since the test sets of IEMOCAP are slightly imbalanced between different emotion categories, in column 6 of Table 1 under the term **Accuracy** we report both Weighted Accuracy (**WACC**) and Unweighted Accuracy (**UACC**).

| Set | # | Input modality | Upstream model | | | Downstream model | | | (Accuracy %) | |
|---|---|---|---|---|---|---|---|---|---|---|
| | | | Input feature | FT | AVG | AGG | Classifier | AVG | WACC | UACC |
| 1.A | 1 | S | W2V2 | yes | no | Mean | Linear | No | 74.09 | 74.56 |
| | 2 | S | huBERT | yes | no | Mean | Linear | No | 72.99 | 73.45 |
| | 3 | S | W2V2 | yes | yes | Mean | Linear | Yes | 76.47 | 76.86 |
| | 4 | S | huBERT | yes | yes | Mean | Linear | Yes | 75.20 | 75.80 |
| 1.B | 5 | S | W2V2 | yes | yes | ECAPA | Linear | yes | 76.58 | 77.07 |
| | 6 | S | huBERT | yes | yes | ECAPA | Linear | yes | 75.56 | 76.78 |
| | 7 | S | huBERT + W2V2 | yes | yes | ECAPA | Linear | yes | **77.36** | **77.76** |
| | 8 | S | huBERT & W2V2 | yes | yes | ECAPA | Linear | yes | 77.04 | 77.52 |
| 2 | 9 | S | Fbank | no | no | ECAPA | Linear | yes | 56.52 | 57.60 |
| | 10 | T | BERT | yes | yes | ECAPA | Linear | yes | 69.34 | 70.07 |
| | 11 | S + T | Fbank & BERT | yes | yes | ECAPA | Linear | yes | 70.56 | 71.46 |

Table 1: Selected experiments to evaluate our SER model: Sets 1.A and 1.B. Baseline experiments: Set 2.

Finally, in column 1 of Table 1 under the term **SET** we have: in (**1.A**) the subset of experiments from Figure 3 that use Mean Pooling as Aggregator; in (**1.B**) the subset of experiments from Figure 3 that use ECAPA-TDNN as Aggregator and in (**2**) the baseline experiments described in Figure 4.

### 4.1 Discussions

From the results presented in Table 1 we can highlight the following observations:
- Comparing experiments 1 and 2 with experiments 3 and 4 we can see that the process of averaging both Upstream models and Downstream models can give WACC improvements of 2.38% and 2.21%, respectively;
- Comparing experiments 5 and 6 with experiments 3 and 4 we can see that the use of ECAPA-TDNN as feature aggregator slightly outperforms mean average pooling;
- Comparing experiment 7 with experiment 8 we can see that early fusion combination of huBERT and Wav2Vec 2.0 features outperforms later fusion;
- Comparing experiment 9 with any experiment from sets 1.A and 1.B we can see that all experiments using either Wav2vec 2.0 or huBERT or both outperforms the standard Filter bank filter by a very large margin.
- Comparing experiment 10 with any experiment from set 1.B we can see that experiments using either Wav2vec 2.0 or huBERT or both outperforms our fine-tuned and Averaged BERT model (which operate on the Ground Truth transcription) by a margin of around 6%.
- Any experiment from Set 1B outperforms experiment 11 by a margin of around 5%.

Table 2 shows a comparison between the results achieved by our method (experiment 7) against the best baselines found on the literature for 5-fold CV on IEMOCAP. The first 4 baselines use audio-only and the fifth baseline uses audio+text. However, the fifth baseline [1] uses the ground truth transcriptions and in addition to that they also use context dependent text embeddings with a window size of [-3,3].

| # | Method | Modalities | (UACC %) |
|---|---|---|---|
| 1 | Sajjad et al. [23] | Audio | 72.25 |
| 2 | Wang et al. [24] | Audio | 73.30 |
| 3 | Liu et al. [25] | Audio | 70.78 |
| 4 | Zhao et al. [26] | Audio | 71.70 |
| 5 | Wu et al. [6] | Audio + Text | 78.30 |
| | **Ours (exp. 7)** | Audio | **77.76** |

Table 2: SOTA results for SER for the cases of audio-only and audio+text input modalities.

As a summary, we can state that we have reached SOTA results for exp. 7 and to the best of our knowledge, the result achieved by exp. 7 is the best result reported so far for SER using 5-fold CV on the IEMOCAP dataset for the case of speech-only input modality.

### 5. CONCLUSIONS AND FUTURE WORKS

In this work we presented an E2E SER model based on an Upstream + Downstream model paradigm which allow easy use of pretrained, fine-tuned and averaged Upstream models. Several experiments using 5-fold CV IEMOCAP dataset were performed, and we have clearly shown that well designed combinations of carefully fine-tuned and averaged Upstream models and averaged Downstream models can significantly improve the performance of E2E SER models. We believe that these results extend even further the possibility of improving E2E SER models by exploiting huge amount of unlabeled data available for pre-training self-supervised acoustic features.

In future work we intend to extend our proposed Upstream + Downstream SER model to *(i)* support multitask learning and *(ii)* exploit multimodality, such as speech + text modalities and speech + text + visual modalities.